\documentstyle[prl,aps,multicol,epsfig]{revtex}

\input epsf.tex

\newcommand{\be}{\begin{equation}}
\newcommand{\ee}{\end{equation}}
\newcommand{\ba}{\begin{eqnarray}}
\newcommand{\ea}{\end{eqnarray}}
\newcommand{\ban}{\begin{eqnarray*}}
\newcommand{\ean}{\end{eqnarray*}}

\newcommand{\ket}[1]{\mbox{$ | #1 \rangle $}}
\newcommand{\bra}[1]{\mbox{$ \langle #1 | $}}

\newcommand{\si}{\sigma}

\newcommand{\compl}{\begin{picture}(8,8)\put(0,0){C}\put(3,0.3){\line(0,1){7}}\end{picture}}

\newcommand{\one}{\leavevmode\hbox{\small1\normalsize\kern-.33em1}}

\def\tr{\mbox{tr}}

\begin{document}

\title{Coherent pulse implementations of quantum cryptography protocols
resistant to photon number splitting attacks}
\author{Antonio Ac\'{\i}n, Nicolas Gisin
and Valerio Scarani}
\address{
Group of Applied Physics, University of Geneva, 20, rue de
l'Ecole-de-M\'edecine, CH-1211 Geneva 4, Switzerland}
\date{\today}
\maketitle

\begin{abstract}
A new class of quantum cryptography (QC) protocols that are robust
against the most general photon number splitting attacks in a weak
coherent pulse implementation has been recently proposed. In this
article we give a quite exhaustive analysis of several
eavesdropping attacks on these schemes. The eavesdropper (Eve) is
supposed to have unlimited technological power while the honest
parties (Alice and Bob) use present day technology, in particular
an attenuated laser as an approximation of a single-photon source.
They exploit the nonorthogonality of quantum states for decreasing
the information accessible to Eve in the multi-photon pulses
accidentally produced by the imperfect source. An implementation
of some of these protocols using present day technology allow for
a secure key distribution up to distances of $\sim$ 150 km. We
also show that strong-pulse implementations, where a strong pulse
is included as a reference, allow for key distribution robust
against photon number splitting attacks.
\end{abstract}

\begin{multicols}{2}

\section{Introduction}

Quantum cryptography, or more precisely, quantum key distribution
(QKD) followed by the one-time pad, is the only secure way of
transmitting secret information (see \cite{review} for a review).
Its security is not based on some mathematical assumptions, such
as a limited eavesdropper's computational power, but on the laws
of Quantum Mechanics. Because of the Heisenberg's uncertainty
principle, a measurement on a quantum system modifies the system
itself. Thus, Eve's measurement on a quantum state carrying
sender's information produces a change on the state that can be
noticed by Alice and Bob. The security of QKD schemes can also be
understood in terms of the no-cloning theorem \cite{WZ}. Eve
cannot make and keep a perfect copy of the quantum state carrying
the information from Alice to Bob \cite{notecl}.

Most of the known QKD protocols use two-dimensional quantum
states, called qubits, as information carriers, although there
exist alternative proposals using higher dimensional systems,
either finite \cite{CBKG} or infinite \cite{GG}. The information
encoding can be performed by means of any two-dimensional quantum
state, but very often this is done using photons that are sent
through an optical fiber, the quantum channel. Therefore, Alice,
must be able to prepare and send single photons to Bob. The
existence of single-photon sources is then an implicit and crucial
requirement for many of the proposed implementation of the
existent schemes. Since there are no practical single-photon
sources available, Alice generally uses a weak coherent pulse,
$\ket{\mu e^{i\theta}}$, with mean photon number $\mu\ll 1$, as an
approximation of the single-photon pulse. Moreover, since there is
no phase reference outside Alice's lab, the effective state used
for the information encoding is
\begin{equation}\label{coh}
    \rho=\int\frac{d\theta}{2\pi}\ket{\mu e^{i\theta}}\bra{\mu
    e^{i\theta}}=\sum_{n}p(n,\mu)\ket{n}\bra{n},
\end{equation}
with the number $n$ of photons distributed according to a Poisson
statistics of mean $\mu$, $p(n,\mu)=e^{-\mu}\mu^n/n!$
\cite{Molmer,lutkenhaus}. Thus, instead of the ideal one-photon
Fock state, Alice produces a zero-photon state with probability
$p(0,\mu)$, a one-photon state with probability $p(1,\mu)$ and so
on.

Intuitively, the presence of pulses with more than one photon may
deteriorate the security of the protocol, since a perfect copy of
the quantum state is then produced by the imperfect single-photon
source. Indeed, it was shown in \cite{lutkenhaus,BLMS} that the
presence of these multi-photon pulses makes the best-known QKD
protocol, the BB84 scheme \cite{BB84}, insecure if the losses in
the channel become important. Eve can then perform the so-called
photon number splitting (PNS) attack that allows her to get full
information without being detected. This limits the distance up to
which BB84 with weak coherent pulses and lossy optical fibers can
be securely implemented. For typical experimental parameters this
critical distance, $d_c$, is of the order of 50 km. As we will
show below, similar conclusions are valid for weak pulse
implementations of other QKD schemes, such as the B92 \cite {B92}
and the 4+2 protocol \cite{HIGM}.

Recently, new quantum cryptography protocols have been proposed
that are more robust against PNS attacks \cite{us}. The scope of
the present article is to give a detailed security analysis of
these protocols under different eavesdropping scenarios. In the
next section we review the PNS attack for the BB84 scheme, and we
show how the same results and conclusions also apply for the B92
and 4+2 protocols. Then, we discuss QKD implementations including
a strong reference pulse as a first possibility for minimizing the
importance of PNS attacks. Moreover, the results of section
\ref{PNS} give us some insight into the requirements needed for a
QKD protocol to be resistant to PNS attacks in a weak pulse
implementations. The family of investigated protocols is presented
in section \ref{prot}. We will focus on a particular one, that
differs from BB84 only in the classical sifting procedure for
extracting the key. We will consider various possible attacks,
some which do not introduce errors, some which use cloning
machines (which do introduce some errors), and some which are the
combination of both. We briefly discuss the experimental data of
\cite{NJP} in the light of our results, as an example of a QKD
implementation secure against PNS attacks. At the end, we also
explore possible generalizations. The last section summarizes the
main results.

Let us end the introduction with four important remarks about our
results. First, in order to make a fair comparison between all the
analyzed protocols, we take as a reference the BB84 scheme using
$\mu=0.1$, i.e. for all the protocols the raw rate at very large
distances must be the same as in this reference scheme. Second, we
do not consider advantage distillation protocols for secret-key
distillation (see for instance \cite{GW}). Therefore, a protocol
is said secure if and only if the information Alice-Bob is greater
than Eve's information. Indeed, it was shown in \cite{CK} that
secret-key distillation is possible using one-way privacy
amplification whenever
\begin{equation}\label{prampl}
    I_{AB}>\min(I_{AE},I_{BE}) .
\end{equation}
Third, although Eve is supposed to have unlimited technological
power, we assume that she is not able to manipulate Bob's detector
(contrary to what is done in Ref. \cite{BLMS}). This is a crucial
point for our analysis \cite{notedet}. Indeed, let us suppose that
Eve can modify the detector efficiency, $\eta_{det}$, in such a
way that it is equal to one for those instances where she has got
the full information (her attack is successful). If we take
$\eta_{det}=0.1$, our results should be modified (see for instance
Eq. (\ref{qevebb84})) by a factor of ten, which means a factor of
10 dB, or equivalently of 40 km, in all our curves. And four, we
do not take into account coherent attacks, where Eve interacts
with more than one pulse \cite{coher}.

\section{The PNS attack}
\label{PNS}

Any experimental realization using photons of a QKD protocol with
two-dimensional quantum states must ideally be performed with a
single-photon source. Unfortunately, this is a very strong
requirement with present day technology, and one has to design a
way of experimentally approximating the single-photon source. In
spite of the fact that QKD has proven to be unconditionally secure
(see for instance \cite{SP}), this may not be the case any longer
if the technology of the honest parties is not perfect.

In most of the existent implementations, the one-photon pulse is
approximated by a weak coherent pulse $\ket{\mu e^{i\theta}}$. As
said above, and since there is no absolute phase reference, the
state seen by Bob and Eve is given by Eq. (\ref{coh}), an
incoherent mixture of multi-photon states with Poisson
probabilities. Eve can then perform a photon number non-demolition
measurement, keep one of the photons when a multi-photon state is
found, and forward the rest to Bob. Note that Eve's action is not
detected by Bob if he is assumed to have only access to the
average detection rate, and not to the statistics of the photons
he receives. We also assume that Eve is able to control the losses
on the line connecting Alice and Bob (or equivalently she can send
photons to Bob by a lossless line). In this situation, Eve can
perform the so-called PNS attack that, as we show below, limits
the security of many of the known existing protocols.

\subsection{The BB84 protocol}

In the BB84 protocol \cite{BB84}, Alice chooses at random between
two mutually unbiased bases, in which she encodes a classical bit.
Denoting by $\ket{\pm x}$ ($\ket{\pm y}$) the eigenvectors of
$\sigma_x$ ($\sigma_y$) with eigenvalue $\pm 1$, she can encode a
logical 0 into either $\ket{+x}$ or $\ket{+y}$ and a 1 into either
$\ket{-x}$ or $\ket{-y}$. She sends the qubit to Bob, who measures
at random in the $x$ or $y$ basis. Then, they compare the basis
and when they coincide, the bit is accepted. In this way, half of
the symbols are rejected, and, in the absence of perturbations,
they end with a shared secret key. In practical situations, and
due to the presence of errors and possibly a spy, some error
correction and privacy amplification techniques have to be
applied, in order to extract a shorter completely secure key.

Now, let us see how Eve can take advantage of the multi-photon
pulses. Alice sends a pulse with $\mu\ll 1$ coding the classical
bit (say on light polarization). Eve performs the photon number
measurement and when two or more photons are detected, she takes
one and forward the rest to Bob by her lossless line. Eve stores
the photon in a quantum memory and waits until the basis
reconciliation. Once the basis is announced, she has only to
distinguish between two orthogonal states, which can be done
deterministically. Thus, for all the multi-photon pulses Eve
obtains all the information about the sent bit. If Alice and Bob
are in principle connected by a lossy line, Eve can block some of
the single-photon pulses, and forward the multi-photon pulses, on
which she can obtain the whole information, by her lossless line.
In this way, Alice and Bob do not notice any change in the
expected raw rate, and Eve remains undetected. When the losses are
such that Eve can block all the single-photon pulses, the protocol
ceases to be secure.

\begin{center}
\begin{figure}
\epsfxsize=8cm \epsfbox{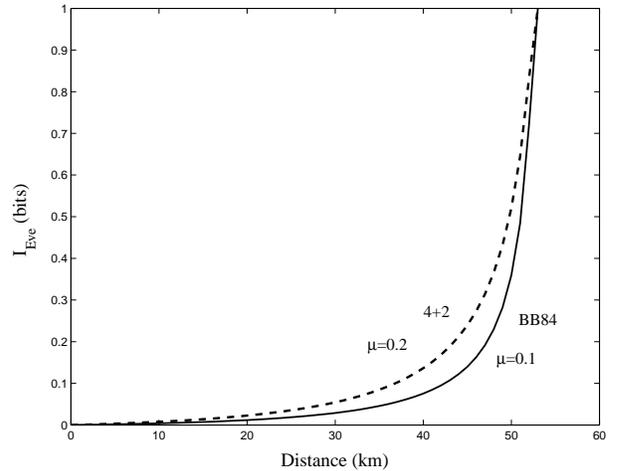} \caption{Eve's information as
a function of the distance for the PNS attacks described in the
text.} \label{pnsatt}
\end{figure}
\end{center}

Denote by $\alpha$ the losses in dB per km on the line. The
expected raw rate at Bob's side is giving by
\begin{equation}\label{rawrate}
    R_{Bob}=\mu\,{10}^{-\delta/10}\ \mbox{
    [photons/pulse]},
\end{equation}
where $\delta=\alpha d$ is the total attenuation in dB of the
quantum channel of length $d$. Eve will apply the PNS attack on a
fraction $1-q$ of the pulses. Since she does not want to be
detected, the raw rate must not change, i.e. she has to choose $q$
in such a way that
\begin{equation}\label{qevebb84}
    R_{Bob}^{PNS}=q\mu+(1-q)R_{BB84}=R_{Bob} ,
\end{equation}
where $R_{BB84}\equiv \sum_{n=2}^\infty p_n(n-1)$ \cite{rates}.
Eve's information is zero when she does nothing, and one for the
PNS attack, i.e. denoting by $S_{BB84}\equiv\sum_{n=2}p_n$,
\begin{equation}\label{ieve}
    I_{eve}(q)=\frac{(1-q)S_{BB84}}{q+(1-q)S_{BB84}} .
\end{equation}
If the losses are such that $q$ can be zero in Eq.
(\ref{qevebb84}) (all the one-photon pulses can be blocked), Eve
gets all the information, without being detected. The critical
attenuation, $\delta_c$, is then given by the condition
$R_{BB84}=R_{Bob}$. In figure \ref{pnsatt} we show the variation
of $I_{eve}$ as a function of $d$ for $\mu=0.1$ and $\alpha=0.25
{\rm dB/km}$ \cite{notealpha}. The critical attenuation in this
case is $\delta_c=13$ dB, and the corresponding distance $d_c=52$
km. Two important points have to be stressed here. First, we do
not claim the optimality of the PNS strategies we consider in this
section from the point of view of Eve's information for losses
lower than $\delta_c$. Indeed, when the losses begin to be
relevant, it is more convenient for Eve to perform the PNS attack
on all the multi-photon pulses and block some of the single-photon
pulses. One can see that this slightly increase $I_{eve}$, but
does not change the critical distance, as defined in this article.
Second, alternative and more conservative definitions of the
critical distance can be proposed. For simplicity, we consider no
perturbations in the absence of Eve, i.e. the information
Alice-Bob, $I_{AB}$, is one. But in realistic situations and due
to the presence of errors (for instance due to detector and
optical noise) this is not true, and the critical distance
corresponds to the point where $I_{eve}=I_{AB}$. If the error rate
is important, this distance may be smaller than the one indicated
here. In any case, for channel attenuations greater than
$\delta_c$, the implementation of the BB84 protocol using weak
coherent pulses is not secure.

One may wonder whether this attack is only possible because the
information is encoded on light polarization. However the same
reasoning is also valid for other encodings such as, for instance,
in the time-bin scheme (see \cite{review}). There the information
is transmitted using the relative phase between two weak coherent
pulses that are sent through the fiber. In principle, the state
leaving Alice's side is $\ket{\phi}=\ket{\mu e^{i\theta}}\ket{\mu
e^{i\theta}e^{i\phi}}$ where $\phi=0,\pi$ ($\phi=\pm\pi/2$)
correspond to $\pm x$ ($\pm y$). But since there is no phase
reference, the effective state seen by Eve and Bob is
\begin{equation}\label{timebin}
    \rho=\int\frac{d\theta}{2\pi}\ket{\phi}\bra{\phi}=
    \sum_{n}p(n,2\mu)\ket{\varphi_n(\phi)}\bra{\varphi_n(\phi)},
\end{equation}
where $p(n,2\mu)$ are Poisson probabilities of mean photon number
$2\mu$ and
\begin{equation}
    \ket{\varphi_n(\phi)}=\sum_{m=0}^n \sqrt{\left(\matrix{n \cr
    m}\right)\frac{1}{2^n}}\,
    e^{i m \phi}\ket{n-m}\ket{m} .
\end{equation}
Note that Bob's state is given by an expression like
(\ref{timebin}) multiplying the mean photon number by the channel
attenuation. It is possible to define a creation and annihilation
operator
\begin{eqnarray}\label{crop}
    a^\dagger(\phi)&=&\frac{a_1^\dagger+
    e^{i\phi}a_2^\dagger}{\sqrt 2} \nonumber\\
    a(\phi)&=&\frac{a_1+e^{-i\phi}a_2}{\sqrt 2} ,
\end{eqnarray}
such that acting on the two-mode vacuum state gives
$a^\dagger(\phi)\ket{0,0}=\ket{\varphi_1(\phi)}$. It is
straightforward to see that
\begin{equation}\label{nphotong}
    \ket{\varphi_n(\phi)}=\frac{(a^\dagger(\phi))^n}{\sqrt{n!}}\ket{0,0} ,
\end{equation}
$[a^\dagger,a]=1$ and
$\bra{\varphi_{n'}(\phi)}\varphi_n(\phi)\rangle=\delta_{n,n'}$.
Thus, the situation is the same as in the previous polarization
encoding scheme \cite{lutkenhaus}. Eve can count the total number
of photons in the two (now temporal) modes, in an analogous way as
in the previous photon number measurement for polarization,
without being noticed by Bob. When ``more than one" photons are
detected, i.e. she projects into $\ket{\varphi_2}$, she stores one
copy of the state in her quantum memory until the basis
reconciliation. Obviously, the equations and critical values in
this case are the same as the ones found above for the
polarization encoding scheme.

\subsection{The B92 protocol}

An alternative QKD scheme is given by the B92 protocol \cite{B92}.
The classical bit is simply encoded by Alice using two
non-orthogonal states, $\ket{\psi_0}$ and $\ket{\psi_1}$ with
$\bra{\psi_0}\psi_1\rangle\neq 0$. Without loosing generality we
will take \cite{comp}
\begin{equation}\label{b92st}
    \ket{\psi_0}=\left(\matrix{\cos\frac{\eta}{2} \cr \sin\frac{\eta}{2}}\right)
    \quad\quad
    \ket{\psi_1}=\left(\matrix{\cos\frac{\eta}{2} \cr -\sin\frac{\eta}{2}}\right) ,
\end{equation}
with $0\leq\eta\leq\pi/2$ and the overlap is
$|\bra{\psi_0}\psi_1\rangle|=\cos\eta$.

Bob has to distinguish between two non-orthogonal quantum states,
and this can only be done with some probability. The measurement
optimizing this probability is defined by the following positive
operators, summing up to one \cite{peres},
\begin{eqnarray}\label{POVM}
    \Pi_0&=&\frac{1}{1+\cos\eta}\,\ket{\psi_1^\bot}\bra{\psi_1^\bot}
    \nonumber\\
    \Pi_1&=&\frac{1}{1+\cos\eta}\,\ket{\psi_0^\bot}\bra{\psi_0^\bot}
    \nonumber\\
    \Pi_?&=&\one-\Pi_0-\Pi_1 ,
\end{eqnarray}
where $\ket{\psi^\bot}$ denotes the state orthogonal to
$\ket{\psi}$. When Bob's measurement outcome is the one associated
to $\Pi_i$, with $i={0,1}$, he knows that the state was
$\ket{\psi_i}$. The probability of obtaining an inconclusive
result is equal to the overlap between the states,
$p_?=\bra{\psi_0}\Pi_?\ket{\psi_0}=\bra{\psi_1}\Pi_?\ket{\psi_1}=\cos\eta$.
Thus, Alice and Bob will accept the sent bit only for those cases
where Bob's measurement gives a conclusive result. The probability
of acceptance is $p_{ok}=1-\cos\eta$, while for the BB84 this
probability is equal to one half. Eve's PNS attack is described in
the following lines.

In a weak pulse encoding scheme, this protocol is clearly
insecure. What Eve can simply do is to perform the same
unambiguous measurement as Bob. When a conclusive result is found,
she knows the state and she prepares a copy of it on Bob's side.
When Eve is not able to determine the state, she blocks the pulse.
Of course, as soon as we have some losses in the channel Alice and
Bob cannot detect the eavesdropping (since they assume that the
absence of signal is due to the losses), and the protocol is
insecure. Note that no quantum memory and lossless line is needed
by Eve in this case.

\subsection{The 4+2 protocol}

\begin{center}
\begin{figure}[c]
\includegraphics[width=4cm]{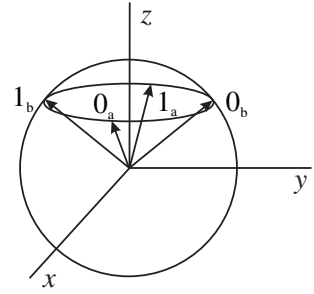}
\caption{Set of states needed for the 4+2 protocol.}
\label{fig42}
\end{figure}
\end{center}

A third QKD protocol was proposed in \cite{HIGM} combining some of
the ideas of the B92 and BB84 schemes. As in the BB84 protocol,
there are four states grouped into two sets
$\{\ket{0_a},\ket{1_a}\},\{\ket{0_b},\ket{1_b}\}$. However, as in
the B92, the states in each set are not orthogonal, their overlaps
being $|\bra{0_a}1_a\rangle|=|\bra{0_b}1_b\rangle|=\cos\eta$. The
situation is depicted in figure \ref{fig42}, the four states lie
on the same parallel of the Bloch sphere. Thus, Alice chooses
randomly in which of the two sets the bit is encoded. Bob performs
at random one of the two POVMs distinguishing the two states of
each set. After basis reconciliation, they determine all the cases
where Bob has applied the correct measurement obtaining a
conclusive result. At first sight, this protocol seems more
resistant against PNS attacks: compared to the BB84 case, Eve can
keep some of the photons but her measurement after the basis
reconciliation may not be conclusive. And compared to the B92
case, she does not know which of the two measurements has to be
applied. However, and due to the particular geometry of the sets
of states, this scheme does not offer any advantage over the two
previous ones. But before describing Eve's attack, let us show how
the three-outcome POVM described by (\ref{POVM}) can be
interpreted as the concatenation of two two-outcome measurements.


The effect of any quantum measurement can be represented by a set
of operators $\{A_i\}$ satisfying $\sum_i A_i A_i^\dagger=\one$.
If the initial state is $\rho$, the probability for any outcome,
say $i$, is
\begin{equation}\label{prmeas}
  p_i=\tr(A_i\rho A_i^\dagger) ,
\end{equation}
and the state is transformed into
\begin{equation}\label{rhmeas}
  \rho_i=\frac{1}{p_i}A_i\rho A_i^\dagger .
\end{equation}
Consider the states (\ref{b92st}). The POVM described by the
operators (\ref{POVM}) can be effectively replaced by a sequence
of two two-outcome measurements. First, one applies a measurement
described by the operators
\begin{eqnarray}\label{filter}
  A_{ok}&\equiv&\frac{1}{\sqrt{1+\cos\eta}}\left(\ket{+x}\bra{\psi_1^\bot}
  +\ket{-x}\bra{\psi_0^\bot}\right) \nonumber\\
  A_?&\equiv&\sqrt{\one-A_{ok}A_{ok}^\dagger} .
\end{eqnarray}
The effect of this first measurement is the following: with
probability $p_{ok}=1-\cos\eta$ the state $\ket{\psi_0}$
($\ket{\psi_1}$) is mapped into $\ket{+x}$ ($\ket{-x}$). This
operation is often called a filtering, and it is equivalent to the
cases where the POVM (\ref{POVM}) gives a conclusive result. When
the outcome $ok$ has been obtained, it is said that the states
have passed the filter. If this is the case, a standard von
Neumann measurement on the $x$ basis suffices for discriminating
between the two states.

Let us come back to the 4+2 protocol and consider the filter for
the states in set $a$, sending these state into the $x$ basis. It
is not difficult to see that the same filter maps the states in
set $b$ into $\ket{\pm y}$. Therefore, a BB84-like situation is
recovered!

It is now easy to design a PNS attack. First, Eve counts the
number of photons. Similar to the B92 case, she applies the
filtering two-outcome measurement when a multi-photon pulse is
obtained. When the result is conclusive, she keeps the resulting
photon in a quantum memory and forwards the rest of photons to
Bob. Then, as in the BB84 case, she waits for the basis
reconciliation, and performs the right von Neumann measurement
allowing her to read the bit. In order to make a fair comparison,
we always impose the same key rate in the absence of Eve as in
BB84 using $\mu=0.1$. In this case this means that we must have
\begin{equation}\label{eqrat}
    \mu_{BB84}=\mu_{4+2}(1-\cos\eta) .
\end{equation}
In a similar way as above for the BB84 case, one can compute Eve's
information for this attack. It almost coincides with the one
found for the BB84 protocol, and the critical distance is again
$\delta_c=52$ km (see figure \ref{pnsatt}). Indeed, the critical
distance turns out to be quite independent of the degree of
non-orthogonality between the states in the 4+2 protocol, if one
imposes the equality of the raw rates (\ref{eqrat}).

The analysis of the 4+2 protocol ends the present section. All the
studied QKD schemes do not guarantee a secure key distillation
when the channel attenuation is around 15 dB. Unfortunately, the
use of non-orthogonal states has not been enough for avoiding
Eve's attacks. The critical distance basically corresponds to the
point where the raw rate on Bob's side can be simulated by the
number of multi-photon pulses leaving Alice's lab.

\subsection{Strong pulse implementations}

The three protocols analyzed in the previous sections are not
robust against PNS attacks in a weak coherent pulse
implementation. Eve exploits the presence of multi-photon pulses
and the losses on the line. Indeed, at the critical distance the
losses are such that she can block the pulse without being noticed
when her attack has not succeeded. A possible way of avoiding this
problem is to send also a strong reference pulse that \textit{must
always be detected} on Bob's side, as in the original B92 proposal
\cite{B92}. In this way, Eve cannot block the pulses without
introducing errors. From the implementation point of view, a new
detector should be added, checking the presence of the strong
pulse. In the following lines we consider these implementations
from the point of view of PNS attacks. We mainly concentrate on
the B92 protocol although, as we will see, the same conclusions
are valid for the other schemes.

The information encoding uses the relative phase between a weak
coherent pulse with respect to a strong reference pulse that is
sent later through the line. Thus, Alice prepares a weak coherent
pulse and a strong pulse sent as a reference,
$\ket{\phi}=\ket{\mu' e^{i\theta}} \ket{\mu e^{i\theta}
e^{i\phi}}$, where $\mu'\gg\mu$ and $\phi=0,\pi$ encodes the
classical bit. For a BB84 scheme, $\phi=0,\pi$ for one of the
basis and $\phi=\pm\pi/2$ for the other. Note that the BB84
implementation with a strong pulse is indeed the 4+2 scheme
\cite{HIGM}. Let us come back to the simplest B92 and denote by
$t$ the ratio between the two intensities $t=\mu/\mu'\ll 1$. The
overlap between the two non-orthogonal states is
$|\bra{0}\pi\rangle|=e^{-2\mu}$, so $\mu< 1$. Bob delays the weak
pulse and makes it interfere with a fraction $t$ of the strong
pulse. Constructive and destructive interference correspond to the
values 0 and $\pi$. The probability of inconclusive result is
$p_?=e^{-2\mu}$ as expected (see \cite{HMGZG} for a practical
implementation of this measurement), and the transmission rate for
small $\mu$ is $\sim 2\mu$ \cite{HIGM}. The detection of the
$1-t\lesssim 1$ fraction of the strong reference pulse by Bob
should allow him to detect Eve's intervention, i.e. none of the
pulses can be blocked. Note that this forces the strong pulse mean
photon number to be significant at Bob's side. However, Eve can
always take advantage of the multi-photon pulses for acquiring
partial information.

Since as usual there is no global phase reference available, the
effective state leaving Alice's lab is
\begin{equation}\label{effstb92}
    \rho=\int\frac{d\theta}{2\pi}\ket{\phi}\bra{\phi}=
    \sum_{n}p(n,\mu+\mu')\ket{\varphi_n(\phi)}\bra{\varphi_n(\phi)},
\end{equation}
where $p(n,\mu+\mu')$ are Poisson probabilities and
\begin{equation}
    \ket{\varphi_n(\phi)}=\sum_{m=0}^n \sqrt{\left(\matrix{n \cr
    m}\right)\frac{t^m}{(1+t)^n}}\,
    e^{i m \phi}\ket{n-m}\ket{m} .
\end{equation}
In a similar way as above, one can define
\begin{eqnarray}\label{cropt}
    a^\dagger(\phi)&=&\frac{1}{\sqrt{1+t}}\left(a_1^\dagger+\sqrt t
    e^{i\phi}a_2^\dagger\right) \nonumber\\
    a(\phi)&=&\frac{1}{\sqrt{1+t}}\left(a_1+\sqrt t e^{-i\phi} a_2\right) ,
\end{eqnarray}
such that acting on the two-mode vacuum state gives
$a^\dagger(\phi)\ket{0,0}=\ket{\varphi_1(\phi)}$. Again, we have
\begin{equation}\label{nphotongt}
    \ket{\varphi_n(\phi)}=\frac{(a^\dagger(\phi))^n}{\sqrt{n!}}\ket{0,0} ,
\end{equation}
$[a^\dagger,a]=1$ and
$\bra{\varphi_{n'}(\phi)}\varphi_n(\phi)\rangle=\delta_{n,n'}$.
Eve can perform a non-demolition measurement for these number
states without being detected by Bob. Indeed, his state is the
same as in Eq. (\ref{effstb92}), just taking into account the
channel attenuation.

Denote the channel losses by $\delta$. Since $\mu'\gg\mu$, Eve's
Poisson distribution is centered around $\mu'$ while Bob's around
$\mu'10^{-\delta/10}$. Moreover the strong pulse must be always
detected by Bob, so we will impose $\mu'10^{-\delta/10}=10$ (at
least), which means that $\mu'=10^{(1+\delta/10)}$. In order to
make a fair comparison with the BB84 scheme using $\mu=0.1$, we
take the same raw rate in the absence of Eve at the critical
distance, which leads to
\begin{equation}\label{ratcomp}
    \frac{\mu_{BB84}}{2}=2\mu_{B92} ,
\end{equation}
and then $\mu_{B92}=0.025$, i.e. $|\bra{0}\pi\rangle|=0.95$, and
$t=10^{-(2+\delta/10)}/4$.

Now, Eve performs the measurement in the $\ket{\varphi_n}$ basis.
Since her Poisson probability is centered around $\mu'$, she
obtains a pulse containing (on average) $\mu'$ photons. On Bob's
side a pulse with ten photons is expected, so Eve keeps
$\ket{\varphi_{\mu'-10}}$ and forwards $\ket{\varphi_{10}}$ to Bob
by her lossless line. Eve's intervention remains unnoticed to Bob.
Eve is now faced with the problem of detecting two states having
an overlap
\begin{equation}\label{b92ov}
    |\langle\varphi_{\mu'-10}(\pi)\ket{\varphi_{\mu'-10}(0)}|=
    \left(\frac{1-t}{1+t}\right)^{\mu'-10}\sim
    \left(\frac{1-t}{1+t}\right)^{\mu'}.
\end{equation}
She applies the measurement maximizing her information
\cite{helstrom}, obtaining
\begin{equation}\label{infest}
   I_{Eve}=I(p_e) ,
\end{equation}
where $I(p)=1+\log_2 p+(1-p)\log_2(1-p)$ is the binary mutual
information (in bits) and $p_e$ is the error probability,
\begin{equation}\label{perror}
    p_e=\frac{1}{2}\left(1-
    \sqrt{1-|\langle\varphi_{\mu'-10}(\pi)
    \ket{\varphi_{\mu'-10}(0)}|^2}\right) .
\end{equation}
It is not hard to compute the limit for Eve's information. For
very large distances, $\mu'\rightarrow\infty$ and then
\begin{equation}
    |\langle\varphi_{\mu'}(\pi)\ket{\varphi_{\mu'}(0)}|=
    \lim_{\mu'\rightarrow\infty}\left(\frac{1-\mu/\mu'}
    {1+\mu/\mu'}\right)^{\mu'}=e^{-2\mu} ,
\end{equation}
i.e. the initial overlap gives the searched limit and $I_{Eve}\sim
0.07$ bits. Thus, for any distance, the protocol is clearly secure
against PNS attacks. The same is valid for the strong pulse
realization of the BB84 protocol, which, as said, is the 4+2
scheme.

Note that strong pulse implementations appear as an intermediate
step in the transition from discrete to continuous variables QKD
schemes using coherent states \cite{GG}. There, a strong reference
pulse, with very large mean photon number $\mu'$, is sent through
the channel with a weaker pulse, containing about hundred photons.
The security comes from the fact that although $\mu$ is not weak,
an infinite range of values is used for the information encoding
(while, for example, we have only two in the B92 case) and Eve is
not able to discriminate the sent state. However, many of the
results presented in this section can be translated to these
protocols, opening the possibility of new eavesdropping attacks
\cite{usprep2}.

Before ending this section let us stress an important point about
strong pulse QKD implementations that was somehow hidden in the
previous discussion. It is important to guarantee a reasonable
photon number for the strong pulse on Bob's side, i.e. the
condition $\mu'10^{-\delta/10}\sim 10$ must be always satisfied.
Therefore, $\mu'$ increases with the distance up to which the key
should be established. Note that $\mu$ is just fixed by the
desired overlap between the two states used in the B92 scheme,
independently of the distance. In the previous lines we took a
quite conservative value, coming from Eq. (\ref{ratcomp}). We can
indeed consider $\mu=1/4$, which gives $|\bra{0}\pi\rangle|=0.6$
and $I_{Eve}\sim 0.5$. This forces $\mu'$ and the ratio $t$ to
increase with the distance, which can lead to problems in the
interferometric arrangement needed for detection. For instance for
a distance of 80 km, that taking as usual $\alpha=0.25$ means 20
dB, we have $\mu'=10^3$ and $t=10^{-4}/4$. However if these
requirements are met, a secure implementation becomes possible
with a key generation rate significantly larger than for the BB84
scheme using $\mu=0.1$.

For the rest of the article however, we will not consider this
type of scenario and we will deal only with implementations using
weak coherent pulses.

\section{QKD protocols resistant to PNS attacks}
\label{prot}

The aim of the present section is to give QKD protocols resistant
to the PNS attack in a weak pulse implementation. From the
previous discussion we can understand some of the basic
requirement for these schemes. We have seen above that the
apparent robustness of the 4+2 protocol was not true due to the
existence of a quantum operation (measurement), represented by
(\ref{filter}), that allows Eve to make pairwise orthogonal the
states in the sets $a$ and $b$. After successfully performing this
operation, she can wait for the basis reconciliation, as in the
BB84 case, and read the information by a von Neumann measurement.
Therefore, what Alice needs is a configuration of sets of states
in which to encode her information such that there does not exist
any quantum operation increasing, even with some probability and
at the same time, the overlap of the states in each set. This is
what a protocol needs to be resistant to the PNS attack.

\begin{center}
\begin{figure}[c]
\includegraphics[width=8.5cm]{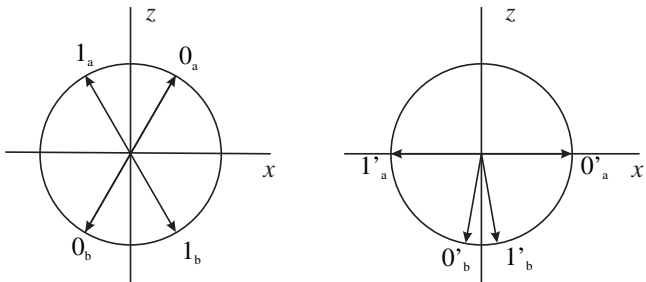}
\caption{States configuration for a QKD protocol robust to PNS
attacks.} \label{figst}
\end{figure}
\end{center}

A simple configuration achieving this property is the same as in
the 4+2, but with one of the two set of states reflected with
respect to the $xy$ plane (see figure \ref{fig42}). But, even
simpler, one can restrict oneself to any plane in the Bloch
sphere, as in the BB84 case. This situation is depicted in figure
\ref{figst}. The general expression for these states is
\begin{eqnarray}\label{stpar}
    \ket{0_a}&=&\left(\matrix{\cos\frac{\eta}{2} \cr
    \sin\frac{\eta}{2}}\right)\quad\quad
    \ket{1_a}=\left(\matrix{\cos\frac{\eta}{2} \cr
    -\sin\frac{\eta}{2}}\right) \nonumber\\
    \ket{0_b}&=&\left(\matrix{\sin\frac{\eta}{2} \cr
    -\cos\frac{\eta}{2}}\right)\quad
    \ket{1_b}=\left(\matrix{\sin\frac{\eta}{2} \cr
    \cos\frac{\eta}{2}}\right)
\end{eqnarray}
After successfully application of the filter that makes orthogonal
the states in set $a$, the overlap between the states in set $b$
has significantly increased. Indeed, it is not difficult to see
that if the outcome of a quantum operation, say $A_i$, makes
orthogonal the states of set $a$, for the same outcome the states
in set $b$ are less orthogonal (see appendix A). So, now Eve has
to consider two different filters, $F_a$ and $F_b$, that make the
states in set $a$ and set $b$ orthogonal, respectively. If she
wants to get the whole information about the bit sent by Alice,
she has to block all the pulses with less than three photons. When
the pulse contains three photons, she applies $F_a$ to the first
one, $F_b$ to the second one, and only when both of them are
conclusive, she forwards the third photon to Bob. It is clear that
the distance where Eve can perform this attack without being
detected is much larger than above. It basically corresponds to
the point where the raw rate is equal to the number of pulses on
Alice's side with more than two photons. There, Eve can simulate
the expected raw rate using only these pulses.


Using this idea, we can design different state configurations. One
of them turns out to be equivalent, at the quantum level, to the
BB84 scheme. The states and the measurements are the same as in
this protocol, the only difference being in the reconciliation
process. But, surprisingly, this variation makes the protocol
significantly more resistant to PNS attacks. The remaining of this
section will be devoted to the detailed security analysis of this
protocol.

\subsection{Four-state protocol}

The configuration of states in figure \ref{figst} allows Alice and
Bob to exchange a key in a secure way for larger distance than for
many of the existing protocols. In the case in which the angle
between the states in each set is $\pi/2$ we recover a BB84-like
state configuration. Nevertheless, note that Alice's bit encoding
has radically changed (see figure \ref{figst}), since orthogonal
states encode the same classical bit.

Suppose as in the standard BB84 that Alice uses as information
carriers the eigenvectors of $\sigma_x$ and $\sigma_y$. Now, the
bit 0 is encoded into $\ket{\pm x}$ and 1 into $\ket{\pm y}$.
Consider the case in which Alice's bit is equal to zero. She
chooses randomly between $\ket{\pm x}$ and sends the state, say
$\ket{+x}$, to Bob. Bob measures randomly in the $x$ or $y$ basis.
After this, Alice starts the reconciliation process announcing the
sent state and one of the two possible states encoding one, for
instance $\{\ket{+x},\ket{+y}\}$. If Bob's measurement was in the
$x$ basis, the result was $+1$ (remember that the sent state was
$\ket{+x}$) and he cannot discriminate between the two
alternatives. If Bob measured in the $y$ basis, for half of the
cases the result was +1 and for the rest -1. In the first case, he
cannot discriminate either, but in the latter, he knows for sure
that the sent bit was not $\ket{+y}$, and accepts the sent bit. At
first sight this is just a simple, and not very useful,
modification of the BB84 protocol. However with these variations
the obtained protocol is much more resistant to Eve's attacks.

Eve is faced with the following problem: after Alice's
announcement she will have to deal with one of four possible sets
of two states,
\begin{eqnarray}\label{setst}
    &&s_1\equiv\{+x,+y\}\quad s_2\equiv\{+y,-x\} \nonumber\\
    &&s_3\equiv\{-x,-y\}\quad s_4\equiv\{-y,+x\} .
\end{eqnarray}
Eve can unambiguously determine the sent state with some
probability for all the pulses of at least three photons. Indeed
she measures in the $x$ and $y$ basis the two first photons, which
allows her to discard two of the possibilities. Then, she applies
to the third photon the measurement discriminating between the two
remaining states. This intuitively shows that this scheme is more
robust against PNS attacks, since only three-photon pulses provide
her with the full information. In the next lines we will extend
these ideas in a more precise way, showing that the distance for a
secure implementation of this protocol is approximately twice the
one for the standard BB84. First we deal with attacks exploiting
the presence of multi-photon pulses without introducing errors on
Bob's side. Then we move to cloning-based attacks, where some
error is allowed, and finally we analyze the combined action of
these two eavesdropping strategies.

\subsection{PNS attacks}

The first type of attacks we consider are of the same type as the
PNS attack for the BB84. Eve uses the multi-photon pulses for
acquiring information. However, her attack cannot be noticed by
the honest parties because of their limited technological powers,
i.e. she must not introduce errors on Bob's side.

As shown above, Eve can determine unambiguously the state sent by
Alice when the pulse contains more than three photons. This is
indeed a general result: unambiguous discrimination between $N$
states of a two-dimensional Hilbert space is only possible when at
least $N-1$ copies of the state are available \cite{cheffles1}. In
this case, the $N$ states $\ket{\psi_i}^{\otimes (N-1)}$ belong to
the symmetric subspace of $(\compl^2)^{\otimes (N-1)}$ of
dimension $N$. Since the $N$ states are always linearly
independent (see appendix B), unambiguous discrimination is
possible. Above we have described a sequence of measurements
allowing unambiguous discrimination between three copies of the
four states $\ket{\pm x},\ket{\pm y}$. The probability of success
is given by the third measurement that discriminates between two
quantum states having an overlap of $1/\sqrt 2$, i.e.
$p_{ok}=1-1/\sqrt 2\sim 0.3$. However, better strategies should be
expected if one acts globally on the three copies of the unknown
state. For instance, one can use the natural generalization of the
POVM described by Eqs. (\ref{POVM}). For any $i=\pm x,\pm y$ one
can define $\ket{\psi_i^\perp}\in(\compl^2)^{\otimes 3}_{sym}$ as
the state orthogonal to the three vectors $\ket{\psi_j}^{\otimes
3}$, with $j\neq i$. Then, the searched measurement is given by
the five positive operators summing up to the identity of
$(\compl^2)^{\otimes 3}_{sym}$, denoted by $\one_{3,sym}$,
\begin{eqnarray}\label{POVMg}
    \Pi_i&=&\frac{2}{3}\ket{\psi_i^\perp}\bra{\psi_i^\perp}
    \nonumber\\
    \Pi_?&=&\one_{3,sym}-\sum_i \Pi_i .
\end{eqnarray}
The probability of having an inconclusive result is, where
$\ket{i^{\,(3)}}=\ket{i}^{\otimes 3}$,
\begin{equation}\label{pincg}
    p_?=\bra{i^{\,(3)}}\Pi_?\ket{i^{\,(3)}}=\frac{1}{2}\, .
\end{equation}
Indeed, this measurement is optimal if we impose that the
probability of conclusive result has to be the same for the four
possibilities to be distinguished. From \cite{cheffles2} we know
that the maximal probability of unambiguous discrimination is
equal to the reciprocal of the maximum eigenvalue of the operator
\begin{equation}\label{chop}
    \frac{1}{4}\sum_{i=\pm x,\pm y} (\ket{\psi_i^\perp}
    \bra{\psi_i^\perp}),
\end{equation}
which gives $p_{ok}=1/2$ \cite{meas}.

\begin{center}
\begin{figure}
\epsfxsize=8cm \epsfbox{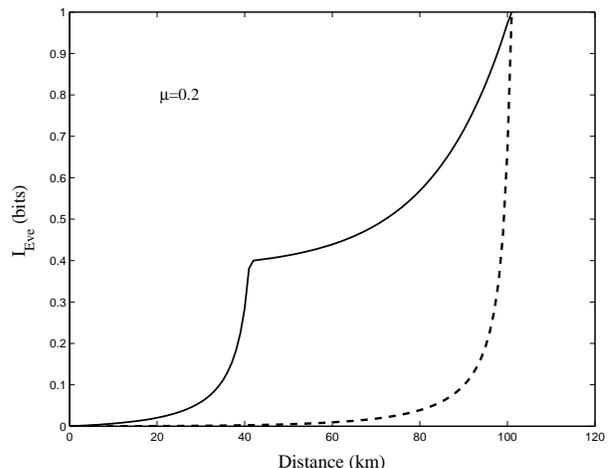} \caption{The figure shows
different eavesdropping attacks that take advantage of the
presence of multi-photon pulses for the new four-state protocol.
The dashed line represents the attack where all pulses with less
than three photons are blocked. Eve can however interpolates
between different attacks as described in the text, depending on
the channel losses. The solid line is Eve's information for this
second possibility.} \label{figiepr}
\end{figure}
\end{center}

Eve's first strategy will consist of counting the number of
photons on each pulse and block those with less than three
photons. When at least three photons are detected, she performs
the previous measurements and blocks all the instances where an
inconclusive result is obtained. When she is able to determine the
state, she can prepare a new copy of it for Bob. Again, if the
channel losses are small, she must apply this strategy on a
fraction of the pulses. The corresponding equations are quite
similar to the ones seen above, and the critical distance
corresponds to the point where
\begin{equation}\label{fincrdist}
    R_{Bob}=p_{ok}^{max}\sum_{n=3}^\infty p_n (n-2) .
\end{equation}
Eve's information is shown in figure \ref{figiepr}, and the
critical distance turns out to be of approximately 100 km. Note
that we take $\mu=0.2$, in order to make a fair comparison with
BB84 using $\mu=0.1$.

It is evident that for small distances, this strategy is quite
inefficient from Eve's point of view. Indeed, for those instances
it is better for her to apply a different PNS attack, that we call
{\sl storing attack}: all single-photon pulses are blocked, while
for all the multi-photon pulses, she keeps one photon in a quantum
memory until the set reconciliation. Then, she has to distinguish
between two non-orthogonal quantum states, say $\ket{+x}$ and
$\ket{+z}$. She will apply the measurement maximizing her
information obtaining (see Eq. \ref{infest})) $I_{eve}\sim 0.4$
and where the error probability is
\begin{equation}
    p_e=\frac{1}{2}\left(1-
    \sqrt{1-|\langle +x\ket{+z}|^2}\right)\sim 0.14 .
\end{equation}
Storing attacks are particularly dangerous as soon as there are
errors in the transmission. If this is the case, the information
$I_{AB}$ is smaller than one and indeed, it may be smaller than
Eve's information (see section \ref{secgenbas} for a more careful
analysis). In a similar way to that described above, depending on
the channel losses, Eve can interpolate between these two attacks.
The corresponding information curves are shown in figure
\ref{figiepr}.

The presence of multi-photon pulses represents a serious drawback,
since Eve can take advantage of them for acquiring information on
the sent bit. Since we do not consider advantage distillation
protocols, the honest parties can extract a key when Eq.
(\ref{prampl}) is satisfied. This means that the secret bit rate
generation, after error correction and privacy amplification, is
\begin{equation}\label{rextr}
    R_{key}=\frac{1}{4}R_{Bob}(1-I_{Eve}) ,
\end{equation}
where $R_{Bob}$ is the raw rate of Eq. (\ref{rawrate}). The $1/4$
term takes into account the set reconciliation process (Bob has to
choose the right measurement and obtain the right outcome), and
the last term comes from the privacy amplification protocol. Note
that we assume for simplicity no errors between Alice and Bob,
$I_{AB}=1$.

\begin{center}
\begin{figure}
\epsfxsize=8cm \epsfbox{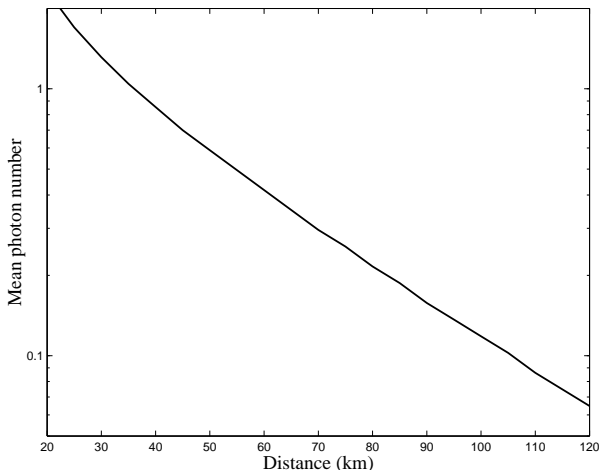} \caption{The figure shows the
mean photon number maximizing the key rate generation, Eq.
(\ref{rextr}), as a function of the distance. For small distances
one cannot take $\mu$ arbitrarily large, since the four states
would become almost orthogonal and Eve could do an
intercept-resend attack without being detected. For large
distances, $\mu$ cannot be arbitrarily small, since the signal
becomes negligible with respect to dark counts and the channel is
completely noisy, $I_{AB}\sim 0$.} \label{muopt}
\end{figure}
\end{center}

There is in principle an obvious way of avoiding the influence of
multi-photon pulses: to decrease the pulse mean photon number.
Nevertheless, this solution may be very inefficient, since the raw
rate, $R_{Bob}$, is approximately proportional to $\mu$.
Therefore, there is a compromise from the point of view of key
generation. Using the same techniques as for figure \ref{figiepr},
for any $\delta$ one can compute the optimal $\mu$ maximizing
$R_{key}$. The corresponding curve is shown in figure \ref{muopt}.
Note that mean photon numbers of the order of 0.2 are indeed
optimal for losses $\sim 20$ dB.

\subsection{Individual attacks using cloning machines}

All the eavesdropping strategies studied up to now take advantage
of the fact that the technological power for the honest parties
has some limitations. In particular, Eve uses the multi-photon
pulses for acquiring information on the sent bit without
introducing any error. Nevertheless, the present protocol must be
also analyzed under the presence of errors, even at the
single-photon level. It may happen that a small amount of error
would allow Eve to gain a large amount of information making the
protocol unpractical. Indeed, these are the attacks Eve would
apply at very short distances, where she cannot block almost any
pulse and almost all the non-empty pulses reaching Bob contain
just one photon.

\begin{center}
\begin{figure}
\epsfxsize=8cm \epsfbox{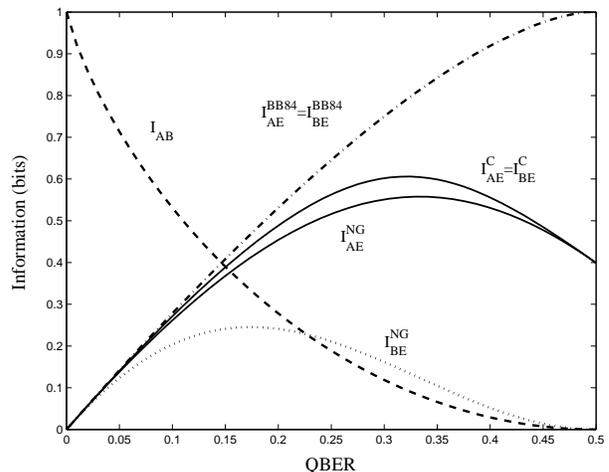} \caption{The figure shows
Alice's and Bob's vs Eve's information for individual attacks
using the cloning machines introduced by Cerf and Niu and
Griffiths. The curve for the standard BB84 scheme is included for
comparison.} \label{ieclon12}
\end{figure}
\end{center}

The optimal individual eavesdropping strategy for this protocol is
unknown. Nevertheless, note that the quantum structure is the same
as for the BB84 scheme, so it seems natural to consider its
robustness against attacks using asymmetric phase covariant
cloning machines \cite{cerf,NG}. These machines, that are briefly
described in appendix C, clone in an optimal way all the states in
the $xy$ plane. Let us stress here that they provide the optimal
eavesdropping for the BB84 protocol \cite{FGGNP}. The action of
these machines in the protocol is depicted in figure
\ref{ieclon12}.

Key distillation using privacy amplification is possible whenever
Eq (\ref{prampl}) is fulfilled. This means that the honest parties
can tolerate an error up to $\sim$ 15$\%$, slightly larger than
the $14.67\%$ for the BB84. There are two facts in these curves
that deserve explanation. First, note that the Cerf cloning
machine \cite{cerf} is clearly more efficient from Eve's point of
view than the Niu-Griffiths one \cite{NG}. Second, note the
surprising decreasing behavior of Eve's information for large
values of the QBER. Both of them are related to the quantum
correlations introduced by each of the cloning machines between
Eve and Bob, and the sifting procedure used in the described
protocol.

Eve waits until the sifting process before doing her measurement.
If, for instance, Alice announces $\ket{+x},\ket{+y}$ and Bob
accepts the symbol, Eve knows that Bob has successfully projected
onto either $\ket{-x}$ or $\ket{-y}$. Then, she modifies her
quantum state according to this information. The fact that Bob has
got a conclusive result (he could discriminate between the two
non-orthogonal states) increases also the distinguishability on
Eve's side because of the quantum correlations. On the one hand,
this justifies why the Cerf cloning machine is more efficient for
eavesdropping. It establishes a stronger correlation between Eve
and Bob, and this helps Eve after the sifting process. On the
other hand, this also explains the decreasing behavior of Eve's
information curves. For very large disturbances, the correlations
between Eve and Bob decreased, and knowing that Bob has obtained a
conclusive result does not help her too much. Thus, it is better
to keep some quantum correlations with Bob, in such a way that his
successful unambiguous discrimination increases the
distinguishability on Eve's side. In the limiting case of
$QBER=0.5$, Eve just takes the state sent by Alice and prepares at
random one of the four possible states for Bob (or in equivalent
terms, she forwards a completely noisy state). Her information is
simply given by Eq. (\ref{infest}) as expected.

\subsection{PNS+cloning attacks}

The eavesdropping strategies analyzed up to now take advantage,
either of the presence of multi-photon pulses (PNS attacks) or of
the errors on Bob's side (cloning attacks). However for losses
such that Eve can simulate the expected rate even if she blocks
all the single-photon pulses, she can combine the two type of
attacks, if she is allowed to introduce some errors. This
basically corresponds to distances $d\gtrsim 40$ km (see figure
\ref{figiepr}). There, Eve counts the number of photons in the
pulse and stops those having one photon. For all the two-photon
pulses, she applies an asymmetric phase covariant $2\rightarrow 3$
cloning machine, and forwards one of the clones to Bob. This
operation introduces errors, depending on the quality of Bob's
clone. In general, for a pulse having $n$ photons, she uses an
$n\rightarrow n+1$ cloning machine, although in this section we
consider only the $2\rightarrow 3$ case, since $p_2$ is
significantly larger than $p_3$. As far as we know this type of
attack has been never considered before. This may explain why the
expression for the phase covariant $n\rightarrow m$ asymmetric
cloning machine is unknown (at least to us). In appendix D we
describe two unitary transformations generalizing, in a
non-optimal way, the asymmetric $1\rightarrow 2$ cloning machines
to the $2\rightarrow 3$ case (see also \cite{usprep,notend}).

\begin{center}
\begin{figure}
\epsfxsize=8cm \epsfbox{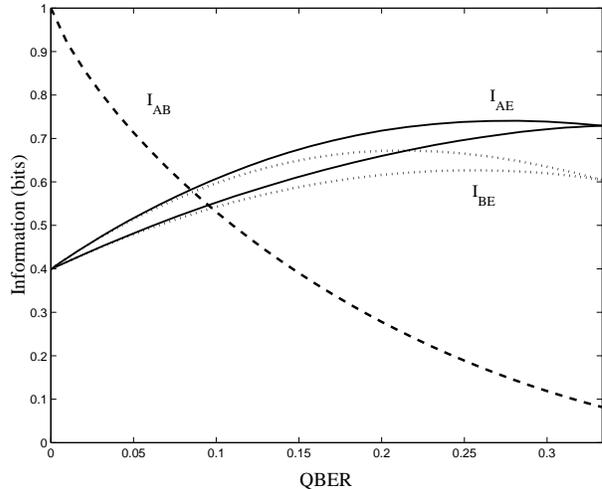} \caption{The figure shows
Alice's and Bob's vs Eve's information for attacks using the
cloning machines described in appendix D. Upper curves correspond
to the cloning machine of Eq. (\ref{clon23ngs}), which is more
powerful from Eve's point of view.} \label{ieclon23}
\end{figure}
\end{center}

Eve counts the number of photons in the pulse. All the
single-photon pulses are blocked, while for those pulses having
two photons she applies one of the $2\rightarrow 3$ cloning
machines shown in appendix D. In this case it is unclear which of
the clone states she has to forward to Bob. It turns out that for
small disturbances, such that Eve's information is smaller than
$I_{AB}$, there is almost no difference between the two cases.
Figure \ref{ieclon23} shows the information Eve can get with this
strategy as a function of the disturbance on Bob's side. We
consider that Bob receives one of the two clones with the same
fidelity, i.e. either the first or the second qubit of Eq.
(\ref{clon23ngs}) or (\ref{clon23c}). Key distillation is possible
using error correction and one-way privacy amplification up to
disturbances of approximately $8.5\%$.

\subsection{The Geneva-Lausanne experiment}

The four-state protocol is at the level of state preparations and
measurements, identical to the BB84 scheme. It only differs in the
sifting process, less efficient in the absence of Eve by a factor
of two on the raw key, but more robust against PNS attacks. Thus,
all the existent experimental implementations of the BB84 protocol
can be thought of as implementations of the new four-state
protocol.

Let us analyze the recent Geneva-Lausanne experiment \cite{NJP},
where a key was distributed over 67 km using the BB84 scheme. The
mean photon number of the pulses used in this experiment was
indeed 0.2 photons/pulse, so all our results directly apply.
According to figure \ref{pnsatt}, the protocol is not secure at
this distance because of the PNS attack even for $\mu=0.1$ (and
BB84 encoding). However this is not the case if one uses the new
protocol. The experimental QBER was approximately $5\%$, where
$4\%$ was due to dark counts on the detector and $1\%$ due to
optical imperfections. As said above, Eve is assumed to have only
access to the optical error. Then $I_{AB}=I(0.05)\sim 0.71$ bits,
while $I_{Eve}$ (see figure \ref{ieclon23}) is clearly smaller
than 0.5. Thus, Alice and Bob can safely distill a key. Note that
even in the more restrictive scenario where Eve can take advantage
of the full error (including the detector noise), her information
is smaller than $I_{AB}$ and the protocol is secure. Therefore,
the existing implementation becomes secure just by changing the
sifting process.

\section{Generalization to more sets}
\label{secgenbas}

\begin{center}
\begin{figure}[c]
\includegraphics[width=4cm]{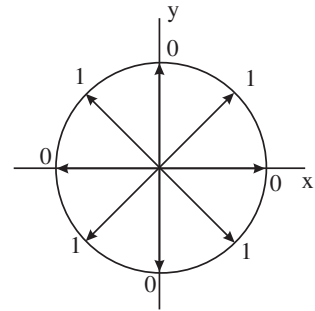}
\caption{Bit encoding in a protocol using four bases.}
\label{genbas}
\end{figure}
\end{center}

The detailed analysis of the four-state protocol has given us
insight into the way of designing QKD protocols resistant to PNS
attacks. The presence of multi-photon pulses is still a problem,
since they open the possibility of unambiguous discrimination or
storing attacks providing Eve full or partial information. But
there is a simple way of improving the robustness of the protocol:
just adding more states for the encoding. A quite natural
generalization of the previous protocol follows this idea and
consists of adding more bases in a plane of the Bloch sphere for
the encoding of the bit, as shown in figure \ref{genbas} for the
case of four bases (eight states). On the one hand more photons
(or copies of the unknown state) are needed for the unambiguous
discrimination to be possible. On the other hand the overlap
between the two announced states decreases, which is also good
against storing attacks. Nevertheless, the key rate decreases
unless we use a larger mean photon number, which increases the
presence of multi-photon pulses, that are dangerous for the
security. Thus, a compromise appears. The aim of this section is
to explore this fact by analyzing the resistance of this
generalized protocols against the two type of attacks mentioned
above: PNS with unambiguous discrimination and storing attacks.

Any protocol is uniquely defined by the number of bases $n_b$ used
for the bit encoding. We will not consider a very large number of
bases, since the protocol would become impractical. In the
previous sections we had $n_b=2$ while $n_b=4$ for the protocol in
figure \ref{genbas}. If Alice wants to send a bit $b$, she chooses
at random between the $n_b$ states encoding $b$ and sends it to
Bob. Bob measures at random in any of the $n_b$ bases. Then, Alice
announces the sent state plus, again randomly, one of the two
neighboring states (encoding $1-b$). Bob accepts the bit when (i)
he has measured in one of the two bases associated to the two
states announced by Alice and (ii) his measurement outcome is
orthogonal to one of these states. Indeed, this allows him to
discard one of the two possibilities and to infer $b$. Thus, Bob
needs to choose the right measurement and obtain the right
outcome, which happens with probability
\begin{equation}\label{prbob}
    p_b=\frac{1}{n_b}\sin^2\left(\frac{\pi}{2n_b}\right) .
\end{equation}
As usual, in order to make a fair comparison, we impose for any
protocol that at very large distances (attenuations) the raw rate
is the same as in the standard BB84 with $\mu=0.1$. This implies
that
\begin{equation}\label{eqratenb}
    \mu(n_b)=\frac{1}{20p_b}=\frac{n_b}
    {20\sin^2\left(\frac{\pi}{2n_b}\right)} .
\end{equation}
Note that for large $n_b$, $\mu(n_b)\sim n_b^3$. This means that
the mean photon number becomes significant when $n_b$ increases
and we are not longer dealing with weak pulses. Therefore, some of
the approximations used above (see \cite{rates}) are not valid.


\begin{center}
\begin{figure}
\epsfxsize=8cm \epsfbox{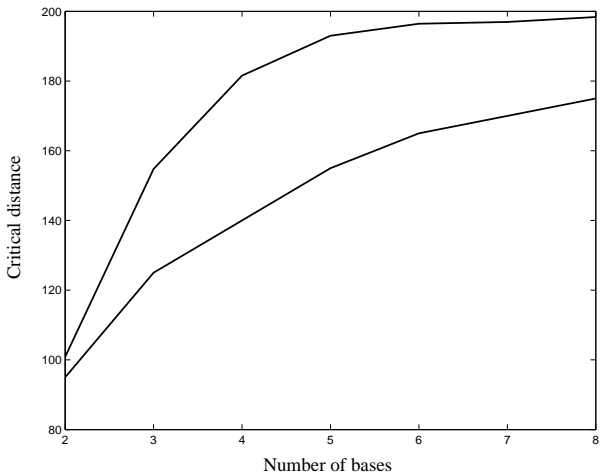} \vspace{.2 cm}\caption{Critical
distance for protocols using $n_b$ bases. Upper curve is given by
PNS attacks using unambiguous discrimination, while the lower
curve corresponds to storing attacks, as explained in the text.
Storing attacks are clearly more efficient from Eve's point of
view.} \label{dcrit}
\end{figure}
\end{center}

Eve has now to discriminate between $2n_b$ one-qubit states, and
this can be done with certainty only when $n_e=2n_b-1$ copies of
the unknown state are available (see \cite{cheffles1} and appendix
B). The maximum probability of success, $p_{ok}$, correspond to
the maximum eigenvalue of the operator \cite{cheffles2}
\begin{equation}\label{chopnb}
    \frac{1}{2n_b}\sum_{k=0}^{n_e} \ket{k^\perp}\bra{k^\perp} .
\end{equation}
Here $\ket{k^\perp}$ denotes the state in $(\compl^2)^{\otimes
n_e}_{sym}$ orthogonal to all $\ket{j}^{\otimes n_e}$, where
$j=0,\ldots,n_e$ but $j\neq k$ and
\begin{equation}
    \ket{k}=\frac{1}{\sqrt 2}\left(\matrix{1 \cr
    e^{i\,k\pi/n_b}}\right) .
\end{equation}
We have numerically calculated these probabilities up to $n_b=8$
and they appear to be given by the formula
$p_{ok}(n_b)=n_b/4^{n_b-1}$, although we do not have an analytical
proof. The critical attenuation $\delta_1$ (in dB) where the
protocol ceases to be secure against this attack has to be such
that Eve can simulate the expected rate by the number of pulses
containing at least $n_e$ photons and giving a conclusive result.
This leads to
\begin{eqnarray}\label{attcrit}
    &&\sum_{n>0}p(n,\mu(n_b)10^{-\delta_1/10})(1-(1-\eta_{det})^n)=
    \nonumber\\
    &&p_{ok}(n_b)\sum_{m\geq n_e} p(m,\mu(n_b))
    (1-(1-\eta_{det})^{(m-n_e+1)}) .
\end{eqnarray}
The corresponding curve is shown in figure \ref{dcrit}.

\begin{center}
\begin{figure}
\epsfxsize=8cm \epsfbox{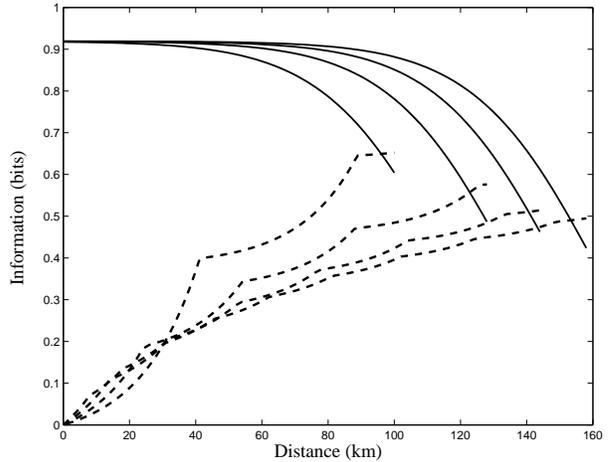} \vspace{.2
cm}\caption{Information curves as a function of the distance for
protocols using $n_b=2,\ldots,5$ bases. Solid lines represent the
information Alice-Bob: at large distances, the signal level is
small compared to dark counts and the QBER becomes important (see
Eq. (\ref{qber})). Dashed lines show Eve's information: at large
distances, she can keep many photons without being detected,
acquiring more information on the sent state. The point where the
two curves cross defines the critical distance where the protocol
is no longer secure.} \label{stattnb}
\end{figure}
\end{center}

There are other attacks, exploiting the presence of multi-photon
pulses, that provide Eve with partial information without
introducing errors. For instance, Eve can count the number of
photons and keep $n_s$, depending on the channel attenuation,
without being detected. She waits until the basis reconciliation
and performs the measurement maximizing her information (see Eq.
(\ref{infest})). These attacks can be very dangerous as soon as we
consider errors on the transmission. We assume that the main
sources of errors are the detector noise, quantified by the
probability $p_d$ of having a dark count, and the optical error
$QBER_{opt}$. The total $QBER$ for a channel attenuation of
$\delta$ is approximately equal to
\begin{equation}\label{qber}
    QBER=\frac{p_d/2}{p_d+\mu(n_b)\eta_{det}10^{-\delta/10}}+QBER_{opt}\, ,
\end{equation}
since half of the dark counts produce a click in the wrong
detector. Thus, for any distance one can compute the amount of
errors and the corresponding $I_{AB}=I(QBER)$. If $I_{Eve}$ is
larger than $I_{AB}$, the protocol is not secure. For any $n_s$,
we can define a critical attenuation such that the honest parties
cannot notice Eve's storing attack. This attenuation corresponds
to the point where
\begin{eqnarray}\label{attcrit2}
    &&\sum_{n>0}p(n,\mu(n_b)10^{-\delta(n_s)/10})(1-(1-\eta_{det})^n)=
    \nonumber\\
    &&\sum_{m\geq n_s} p(m,\mu(n_b))
    (1-(1-\eta_{det})^{(m-n_s)}) .
\end{eqnarray}
For intermediate attenuations (distances), Eve can interpolate
between two attacks, as described above. In this way, we can
compute the two curves $I_{AB}$ and $I_{Eve}$ as a function of the
distance. Figure \ref{stattnb} shows the obtained results, where
we took $\eta_{det}=0.1$, $p_d=10^{-5}$ and $QBER_{opt}=1\%$. The
point where $I_{AB}=I_{Eve}$ provides the critical distance,
$\delta_2$, for this type of attacks. In figure \ref{dcrit} we
plot both the $\delta_1$ and $\delta_2$ as a function of $n_b$. It
is quite plausible that $\min(\delta_1,\delta_2)$ gives a good
estimation for $\delta_c$, the critical distance associated to the
unknown optimal attack. Thus, one can safely conclude that a key
can be established using a reasonable number of bases up to
distances of the order of 150 km \cite{othergen}.

\section{Conclusions}
\label{concl}

Unconditional security of quantum cryptography relies on some
experimental assumptions that are not achievable with present-day
technology. Thus, in a more realistic scenario, the honest parties
have to deal with approximated single-photon sources, noisy
channels, inefficient detectors and so on, while no limitation on
the eavesdropper technology should be assumed. This opens the
possibility for alternative eavesdropping attacks, taking
advantage of Alice and Bob's technological imperfections. Indeed,
using as a reference the BB84 scheme with $\mu=0.1$, all the known
protocols become insecure against PNS attacks for channel losses
of the order of 13 dB.

In this article we show how to construct QKD protocols resistant
against PNS attacks up to channel losses of 40 dB. There are two
possibilities for that: (i) to exploit the non-orthogonality of
quantum states in a different way, as in the presented four-state
protocol or (ii) to include a strong reference pulse that must be
always detected by Bob. Both possibilities seem achievable with
current technology. In the first case, already existent
implementations of the BB84 protocol \cite{NJP} provide an
experimental demonstration of QKD secure against PNS attacks, when
the alternative sifting process is applied. The second possibility
shows a connection between discrete and continuous variables QKD
schemes that deserves further investigation.

\section{Acknowledgements}

We thank Nicolas Cerf, Daniel Collins, Norbert L\"utkenhaus and
Gr\'egoire Ribordy for helpful discussion. We acknowledge
financial supports by the Swiss OFES and NSF within the European
project RESQ (IST-2001-37559) and the national center "Quantum
Photonics".

\section*{Appendix A}

In this appendix we show that the overlap between all the states
in figure \ref{figst} cannot be decreased by the same quantum
operation. Using the parametrization of Eq. (\ref{stpar}), one can
see that
\begin{eqnarray}\label{strel}
    \ket{0_b}&=&c\ket{0_a}+c'\ket{1_a} \nonumber\\
    \ket{1_b}&=&c'\ket{0_a}+c\ket{1_a} ,
\end{eqnarray}
where
\begin{equation}\label{stcoeff}
    c=-\frac{\cos\eta}{\sin\eta}\quad\quad c'=\frac{1}{\sin\eta} .
\end{equation}
Now, consider a quantum operation, $M$, mapping with some
probability $p_a$ the states in set $a$ into some new states,
$\ket{0'_a}$ and $\ket{1'_a}$, such that $\langle
0'_a\ket{1'_a}=0$. This means that
\begin{equation}
    M\ket{i_a}=\frac{1}{\sqrt p_a}\ket{i_a'} ,
\end{equation}
where $i=0,1$. Because of the linearity of Quantum Mechanics, the
states in set $b$ will be mapped into
\begin{eqnarray}\label{newstb}
    \ket{0'_b}&=&\frac{1}{\sqrt{p_b}}(c\ket{0'_a}+c'\ket{1'_a})
    \nonumber\\
    \ket{1'_b}&=&\frac{1}{\sqrt{p_b}}(c'\ket{0'_a}+c\ket{1'_a}) ,
\end{eqnarray}
with probability
\begin{equation}
    p_b=\frac{1+\cos^2\eta}{\sin^2\eta}\frac{1}{p_a} .
\end{equation}
Their overlap is
\begin{equation}
    |\langle
    0'_b\ket{1'_b}|=\frac{2\cos\eta}{1+\cos^2\eta}\geq\cos\eta ,
\end{equation}
i.e. the states in set $b$ become less distinguishable.

\section*{Appendix B}

In this appendix we will show that $N-1$ copies of $N$ one-qubit
state are always linearly independent (see also \cite{cheffles1}).
Consider $N-1$ copies of $N-1$ general states of one qubit,
$\ket{\psi_i}$ with $i=1,\ldots,N-1$. They belong to the symmetric
subspace $(\compl^2)^{\otimes(N-1)}_{sym}$ of dimension $N$. Our
aim now is to add a new state and to see when this state can be
written as a linear combination of the previous ones. In other
terms, we want to find a state $\ket{\psi_N}\in\compl^2$ such that
the determinant of the $N\times N$ matrix
\begin{equation}\label{determ}
    \left(\matrix{\ket{\psi_1}^{\otimes(N-1)}\cdots
    \ket{\psi_{N-1}}^{\otimes(N-1)}\ket{\psi_N}^{\otimes(N-1)}}\right)
\end{equation}
is zero. Note that the norm of the state does not play any role,
so we can write
\begin{equation}
    \ket{\psi_N}=\left(\matrix{1 \cr x}\right) ,
\end{equation}
where $x$ is an unbounded complex number. Condition (\ref{determ})
then gives an $N-1$ degree polynomial equation on $x$. There are
$N-1$ solutions, that correspond to the $N-1$ trivial cases
$\ket{\psi_N}=\ket{\psi_i}$ for $i=1,\ldots,N-1$. Thus, $N-1$
copies of any $N$ different one-qubit state are always linearly
independent.

\section*{Appendix C}

In this appendix we briefly describe the asymmetric phase
covariant cloning machines introduced in \cite{cerf,NG}. These
machine clone with maximal fidelity all the states that lie in the
$xy$ plane. At first sight, their only difference is that the one
in \cite{cerf} uses as an input state a two-qubit reference state
plus the state to be cloned, while for the second one qubit
suffices as ancillary system.

Consider an input state to be cloned, and a one-qubit ancillary
system in a reference state, say $\ket{0}$. The Niu-Griffiths
cloning machine \cite{NG} is defined by the following unitary
transformation
\begin{eqnarray}\label{clng}
    U_{12}^{NG}\ket{00}_{12}&=&\ket{00} \nonumber\\
    U_{12}^{NG}\ket{10}_{12}&=&\cos\gamma\ket{10}+\sin\gamma\ket{01} ,
\end{eqnarray}
with $0\leq\gamma\leq\pi/2$. From the definition it is evident
that this transformation does not affect in the same way the two
poles $\ket{\pm z}$ of the Bloch sphere. Nevertheless, this is not
the case for those state lying in the $xy$ plane, i.e.
$\ket{\vartheta}=(\ket{0}+e^{i\vartheta}\ket{1})/\sqrt 2$. The
searched clones are the mixed local states resulting from tracing
either the first or the second qubit on the state resulting from
the application of Eq. (\ref{clng}),
\begin{equation}\label{stclng}
    \rho_i=\tr_{2-i}(\Pi_{NG}(\vartheta)) ,
\end{equation}
where $i=1,2$ and $\Pi_{NG}(\vartheta)$ is the projector onto
$U_{NG}\ket{\vartheta}\ket{0}$. One can easily see that
$\forall\,\vartheta$
\begin{eqnarray}\label{stclng2}
    \rho_1=\cos\gamma\ket{\vartheta}\bra{\vartheta}+
    (1-\cos\gamma)\,\frac{\one}{2} \nonumber\\
    \rho_2=\sin\gamma\ket{\vartheta}\bra{\vartheta}+
    (1-\sin\gamma)\,\frac{\one}{2} .
\end{eqnarray}
Then, the corresponding clone fidelities, defined as
$F_i=\bra{\vartheta}\rho_i\ket{\vartheta}$, are $(1+\cos\gamma)/2$
and $(1+\sin\gamma)/2$. The larger the fidelity for the first
clone, the smaller for the second. Equality is achieved when
$\cos\gamma=\sin\gamma$, and then $F_1=F_2=(1+1/\sqrt 2)/2$.

The second type of cloning machine we consider are those
introduced in \cite{cerf}. There, two qubits are used as the
ancillary system, and the unitary transformation is, for any input
state $\ket{\psi}\in\compl^2$,
\begin{eqnarray}\label{clcerf}
    U_{12}^{C}\ket{\psi}\ket{00}&=&F\ket{\psi}\ket{\Phi^+}+
    (1-F)\si_z\ket{\psi}\ket{\Phi^-}+\nonumber\\
    &&\sqrt{F(1-F)}\left(\si_x\ket{\psi}\ket{\Psi^+}+
    i\si_y\ket{\psi}\ket{\Psi^-}\right) ,
\end{eqnarray}
where
\begin{eqnarray}\label{Bellbasis}
    \ket{\Phi^\pm}=\frac{1}{\sqrt
    2}\left(\ket{00}\pm\ket{11}\right) \nonumber\\
    \ket{\Psi^\pm}=\frac{1}{\sqrt
    2}\left(\ket{01}\pm\ket{10}\right)
\end{eqnarray}
define the standard Bell basis. It is not difficult to see that
the local state in the first two qubits is the same as in Eq.
(\ref{stclng}), if one takes $F=(1+\cos\gamma)/2$.

Eve can use these transformations in order to obtain some
information about the sent bit. She clones the state sent by
Alice, and she forwards the first clone to Bob and keeps the
second. Obviously there is a compromise between the quality of the
two clones: the better Eve's clone the worse Bob's state. Or in
other words, the more the information intercepted by Eve, the more
the errors on Bob's side, that allow the honest parties to detect
Eve's intervention. As seen above, the two machines are in many
senses equivalent (especially as far as for the cloning fidelities
are concerned). However the two attacks differ in the amount of
correlations Eve establishes with Bob. This fact is going to be
very important for the type of protocols analyzed in this work.

\section*{Appendix D}

In this appendix we give two different unitary transformations
that somehow generalizes the $1\rightarrow 2$ asymmetric cloning
machines to the $2\rightarrow 3$ case. The complete description of
these machines will be given elsewhere \cite{usprep}.

\begin{center}
\begin{figure}
\epsfxsize=8cm \epsfbox{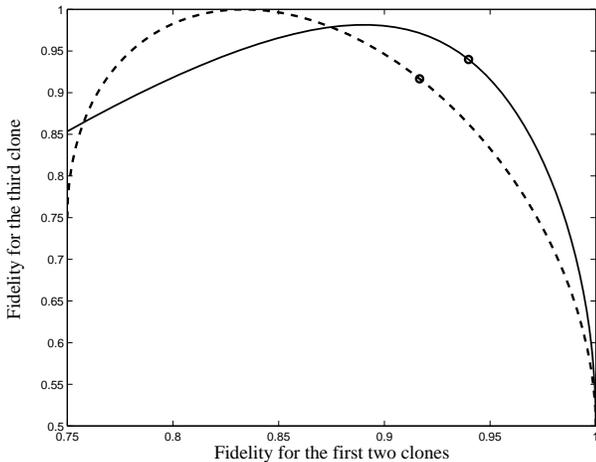} \caption{Cloning fidelities
for the $2\rightarrow 3$ cloning machines defined by Eqs.
(\ref{clon23ng}) (solid line) and (\ref{clon23c}) (dashed line).
The circles correspond to the points where the cloning fidelities
are equal.} \label{clonfid}
\end{figure}
\end{center}

The first machine is mainly inspired by Niu-Griffiths
construction. The initial input state corresponds to two copies of
an unknown one-qubit state, $\ket{\psi}^{\otimes
2}\in(\compl^2\otimes\compl^2)_{sym}$. Using a two-dimensional
ancillary system, say in state $\ket{0}$, one can define the
unitary operation
\begin{eqnarray}\label{clon23ng}
    U_{23}^{NG}\ket{00}\ket{0}&=&\ket{000} \nonumber\\
    U_{23}^{NG}\ket{\Phi^+}\ket{0}&=&\frac{\cos\gamma(\ket{010}+
    \ket{100})+\sin\gamma\ket{001}}{\sqrt{1+\cos^2\gamma}} \nonumber\\
    U_{23}^{NG}\ket{11}\ket{0}&=&\frac{\cos\gamma\ket{110}+
    \sin\gamma(\ket{011}+\ket{101})}{\sqrt{1+\sin^2\gamma}} .
\end{eqnarray}
As in the $1\rightarrow 2$ case, this machine has not the same
effect on the states $\ket{0}$ and $\ket{1}$. After some lengthy
algebra one can see that all the states $\ket{\psi}$ in the $xy$
plane are cloned with the same fidelity. The local state of each
of the three qubits is a combination of the identity with the
initial pure state as expected, the fidelities being (see also
figure \ref{clonfid})
\begin{eqnarray}\label{clfidng}
    F_1^{NG}&=&F_2^{NG}=\frac{1}{2}+
    \frac{\cos\gamma}{2\sqrt{3+\cos(2\gamma)}}+
    \frac{1}{\sqrt{17-\cos(4\gamma)}} \nonumber\\
    F_3^{NG}&=&\frac{1}{2}+
    \frac{\sin\gamma}{2\sqrt{3+\cos(2\gamma)}}+
    \frac{\sin(2\gamma)}{\sqrt{17-\cos(4\gamma)}}
    .
\end{eqnarray}
Note that when $\gamma=\pi/4$, $F_1^{NG}=F_3^{NG}=(6+2\sqrt
2+\sqrt 6)/12\sim 0.94$, slightly larger than the fidelity of the
$2\rightarrow 3$ universal symmetric cloning of \cite{GM}. It has
to be stressed that the fidelity for the third clone never reaches
the value of one, contrary to what happens for the $1\rightarrow
2$ case. As we learnt from the analysis of the individual attacks,
in the type of protocols we analyze it is more convenient for Eve
to introduce an extra ancillary system in such a way that she is
better correlated to Bob's result. This can be done introducing an
ancillary system on Eve's side, such that the action on the states
of the computational basis is symmetrized. Note that in the
$1\rightarrow 2$ case this procedure allows to pass from the
Niu-Griffiths to the Cerf cloning machine. The resulting machine
can be expressed as,
\begin{equation}\label{clon23ngs}
    U_{23}^{NGs}\ket{s}\ket{00}=(U_{23}^{NG}\ket{s}\ket{0})
    \ket{0}+(\tilde U_{23}^{NG}\ket{s}\ket{0})\ket{1} ,
\end{equation}
where $\ket{s}=\ket{00},\ket{\Phi^+},\ket{11}$ and $\tilde
U_{23}^{NG}$ has the same form as $U_{23}^{NG}$ but interchanging
zeros and ones. The cloning fidelities are again equal to Eq.
(\ref{clfidng}).

The second machine we consider is based on the Cerf construction
\cite{cerf}. As an input state we have two qubits of an unknown
one-qubit state plus a two-qubit ancillary system. Then, we define
the following unitary operation,
\begin{eqnarray}\label{clon23c}
    U_{23}^{C}\ket{\psi}^{\otimes 2}\ket{00}&=&v\ket{\psi}^{\otimes 2}
    \ket{\Phi^+}+
    x(\tilde\si_z\ket{\psi}^{\otimes 2}\ket{\Phi^-}+\nonumber\\
    &&\tilde\si_x\ket{\psi}^{\otimes 2}\ket{\Psi^+}+
    i\tilde\si_y\ket{\psi}^{\otimes 2}\ket{\Psi^-}) ,
\end{eqnarray}
where, for $k=x,y,z$,
\begin{equation}
    \tilde\si_k=\si_k\otimes\one+\one\otimes\si_k ,
\end{equation}
and $v^2+8x^2=1$. One can see that for any input state in the
Bloch sphere, the local state of the first two qubits are two
identical clones with fidelity $F_1^C=F_2^C=1-2x^2$, while in the
third qubit we have another clone with fidelity
$F_3^C=1-(v-3x)^2/2$. Thus, the machine (\ref{clon23c}) is an
asymmetric universal cloning machine, i.e. not phase covariant.
Indeed, at the point where the three fidelities are equal, we
recover the $2\rightarrow 3$ cloning fidelity of \cite{GM}
$F_1^C=F_3^C=11/12$ (see also figure \ref{clonfid}). Note also
that in this case, $F_3^C$ can be equal to one. Moreover, there
are some points where, for a given fidelity for the first two
clones, the fidelity for the third one is larger using this
cloning machine than for the phase covariant machine of Eq.
(\ref{clon23ng}). This shows that the latter is not the optimal
phase covariant asymmetric $2\rightarrow 3$ cloning machine. One
is tempted to generalize Cerf construction in a direct way,
defining a phase covariant machine by changing the coefficient of
one of the error terms in (\ref{clon23c}). However, the resulting
operation is not unitary \cite{usprep}. Therefore, we can only
propose two possible asymmetric phase covariant machines, although
we know that they are not optimal. Nevertheless, it is quite
reasonable to suppose that the increase on Eve's information will
not be very significant when using the, at present unknown,
optimal one \cite{notend}.

\end{multicols}


\begin{thebibliography}{99}
\bibitem{notecl}
It is a well-known fact that the existence of a perfect quantum
cloning machine would allow to beat Heisenberg's uncertainty
principle.
\bibitem{notedet}
One may think, for instance, that Eve is able to shift the signals
she wants to be detected into a wavelength region of higher
detector efficiency. But this can be simply avoided by putting a
narrow filter before the detector. Or she can send pulses with
large mean photon number whenever the pulse should be detected,
but this produces a significant increase of the double counts when
Bob chooses the wrong measurement. Thus, we believe the only
possibility left is to give Eve access to Bob's detector. But
then, what prevents her to control also Bob's computer, i.e. to
know Bob's bit?
\bibitem{rates}
More precisely, what Eve has to reproduce on Bob's side is the
number of non-empty pulses, $1-p(\mu,0)$. However, for the small
mean photon numbers considered, this is approximately equal to the
mean rate. Note also that we take for simplicity the detector
effciency equal to $\eta_{det}=1$. Indeed, for small $\mu$, this
factor affects in approximately the same way all the pulses,
independently on whether Eve performs the PNS attack, since it is
assumed that Eve has no control on Bob's detector. Or, in other
words, $\eta_{det}$ appears on both sides of Eq. (\ref{qevebb84}).
Therefore, the critical distances do not depend on the detector
efficiency.
\bibitem{notealpha}
We take $\alpha=0.25$ dB/km for all the figures in this paper.
\bibitem{comp}
In all the article, we take as the computational basis, $\ket{0}$
and $\ket{1}$, the eigenvectors of $\si_z$ with eigenvalues $\pm
1$.
\bibitem{meas}
The optimal generalized measurement of Eq. (\ref{POVMg}) in
$(\compl^2)^{\otimes 3}_{sym}$ can indeed be seen as a von Neumann
measurement in $(\compl^2)^{\otimes 3}$ \cite{us}. It is very
plausible that there is no measurement using just linear optics
reaching the optimal probability of unambiguous discrimination.
\bibitem{othergen}
There may be other generalizations of the initial four-state
protocol. For instance, for $n_b=4$, one can consider a bit
encoding different from the one in figure \ref{genbas}. The bit
$b=0$ can be associated to the states
$\ket{0},\ket{\pi/4},\ket{\pi},\ket{5\pi/4}$, where
$\ket{\vartheta}=(\ket{0}+e^{i\vartheta}\ket{1})/\sqrt 2$, and
$b=1$ to the other four states. In the sifting process, Alice
always announces two states having an overlap of $1/\sqrt{2}$, as
in the initial four-state protocol. In this way, (i) the
discrimination on Bob's side is more robust against imperfect
measurement apparatus and (ii) the probability of accepting a bit
is greater, $p_b=1/(2n_b)$, and then $\mu(n_b)$ only increases
linearly with $n_b$. Since the mean photon number does not need to
be very large for having the same key rate generation, the number
of multi-photon pulses at a given distance is smaller, and the
protocol is more secure against PNS attacks using unambiguous
discrimination. However, when one considers storing attacks, the
protocol is not efficient. Indeed, Eve can always keep some
photons, $n_s$, without being detected and wait for Alice's
announcement. Then, she has to distinguish between $n_s$ copies of
two states with overlap $1/\sqrt 2$. While $n_s$ increases with
the number of bases, the overlap is independent of $n_b$.
Therefore, to increase $n_b$ does not provide any advantage to the
honest parties when they use this alternative encoding.
\end{thebibliography}
\end{document}